\documentclass[seceq]{ptptex}
\usepackage{amsmath,amscd}

\notypesetlogo
\title{%
Gauge Freedom in the Path Integral Formalism}

\author{%
Seiji {\scshape Sakoda}%
}

\inst{%
Department of Applied Physics,
National Defense Academy,\\
Yokosuka 239-8686, Japan
}

\abst{We investigate 't Hooft's technique of changing the gauge
  parameter of the linear covariant gauge from the point of view of
  the path integral with respect to the gauge freedom. Extension of
  the degrees of freedom allows us to formulate a system with extended
  gauge symmetry. The gauge fixing for this extended symmetry yields
  the 't~Hooft averaging as a path integral over the additional
  degrees of freedom. Another gauge fixing is found as a non-abelian
  analogue of the type II gaugeon formalism of Yokoyama and Kubo. In
  this connection, the 't~Hooft average can be viewed as the analogue
  of the type I gaugeon formalism.  As a result, we obtain gauge
  covariant formulations of non-abelian gauge theories, which allow us
  to understand 't~Hoot's technique also from the canonical
  fromalism.}
\begin{document}

\maketitle

\section{Introduction}
\label{sec:intro}
In the quantization of non-abelian gauge fields, the path
integral~(PI) formalism is a powerful and efficient tool. In
particular, in this formalism changing gauge fixing conditions can be
carried out in a very simple manner. For instance, the change of
gauges from the physical Coulomb gauge to the Landau gauge first
presented by Faddeev and Popov~(FP)~\cite{FP} and its
generalization~\cite{tHooft} formulated by 't~Hooft to any linear
covariant gauge cannot be treated in the canonical quantization,
because the latter in a specific gauge requires its own Hilbert space
and there are no gauge transformations which generate transitions
between different Hilbert spaces.  Therefore, to clarify the
equivalence of two or more theories in different gauges in the
canonical formalism, we must compare the results obtained from each of
the quantum theories.  In the path integral formalism, by contrast, we
can transform between different gauges in a relatively simple manner,
by inserting an identity and carrying out a change of variables.

As far as the canonical formulation of quantum electrodynamics is
concerned, there is an elegant prescription, the gaugeon formalism of
Yokoyama and Kubo,~\cite{gaugeon} \ to carry out the change of the
gauge parameter $\alpha$ in linear covariant gauges. By taking the
concept of the \textit{form invariance} of the Lagrangian in this
formalism as a principle, generalizations of Yang-Mills theory have
also been explored.~\cite{gaugeonYM,YTM} \ Unfortunately, however, the
concept of the form invariance does not seem to be compatible with the
gauge condition of linear covariant gauges for the case of non-abelian
gauge theories, and for this reason, the gauge parameter $\alpha$
cannot be controlled by means of the $q$-number gauge transformation,
which is a global transformation defined on the extension of the
degrees of freedom for the gaugeon field and its associated field; a
very complicated Lagrangian is necessary in order to implement the
form invariance for the case of a non-abelian theory, and the
$q$-number gauge transformation then generates changes in parameters
of the Lagrangian in a manner different from that in the case of the
abelian theory. Contrastingly, in the case of the path integral
formalism, 't~Hooft formulated a very simple method~\cite{tHooft} of
changing the gauge parameter $\alpha$. This method consists of
averaging with respect to an unknown $c$-number function that enters
the PI when the condition of the Landau gauge is slightly modified.
In comparison with the complexity of the gaugeon formalism of
Yang-Mills theory, the simplicity of this method represents a great
advantage of the path integral formalism, and due to this convenience
the path integral formalism is more practical than the canonical
formalism in formulating a quantization of the Yang-Mills theory. In
this paper, we investigate the meaning of such an advantage of the
path integral method, first considering the 't~Hooft
averaging~\cite{tHooft} from our own point of view. (Throughout this
paper, we refer to the averaging over a $c$-number function in the
path integral formalism performed to change a gauge condition in this
way as ``'t~Hooft averaging.'') Here, similarly to Yokoyama's gaugeon
formalism,~\cite{gaugeon} \ a pair of fields plays a role in finding a
new interpretation of this technique. Our proposal here is to view the
't~Hooft averaging as a PI obtained through gauge fixing for an
extended gauge invariance.~\cite{exgauge} \ As a technical tool for
changing the gauge condition, the 't~Hooft averaging shares much with
the gaugeon formalism. If our interpretation of the 't~Hooft average
is correct, it will allow us to treat such an extended system in terms
of a canonical quantization similar to that in the case of the gaugeon
formalism.  As mentioned above, however, a gaugeon formalism suitable
for handling the linear covariant gauge is known only for the abelian
gauge theory.  In this paper, we attempt to realize the above-stated
goal for non-abelian gauge theories.

We first introduce a change of variable for the $c$-number function in
the 't~Hooft average to reformulate the PI with extended degrees of
freedom in the next section. Then, in \S~3, an extended gauge
transformation is introduced to make the system invariant under this
extended transformation. Using the usual FP trick for such an
invariant PI, we obtain the 't~Hooft averaging as a result of the
gauge fixing. Because we utilize the FP trick for the gauge fixing, a
PI possessing invariance under BRST transformations~\cite{BRST} is
naturally obtained. It is shown that the extended BRST transformation
consists of two anti-commuting generators. On the basis of these
generators, we carry out analyses of the BRST invariance and the gauge
structure of the total Hilbert space for the extended system.  In the
derivation of the 't~Hooft average discussed above, however, we still
need an averaging with respect to another $c$-number field. To avoid
this, we formulate another prescription for the gauge fixing. The
difference between the two gauge fixing procedures is found to be
similar to that between the type I and type II gaugeon formalisms of
Yokoyama and Kubo.~\cite{gaugeon} \ Despite the difference in their
Lagrangians, the structure of the Hilbert space and the BRST
invariance of the system obtained with this second method are quite
similar to those of the first method. In this way, we obtain
non-abelian analogues of type I and type II gaugeon formalism. This is
the contents of \S~3. Comparison of our method for the non-abelian
case with the derivation of the PI presented by Koseki, Sato and
Endo~\cite{gaugeonKSE} for the BRST invariant version of Yokoyama's
gaugeon formalism is made in \S~4. We also explain there how to change
the gauge parameter in our formalism, confirming that our method can
be regarded as a non-abelian generalization of the gaugeon formalism
suitable for the linear covariant gauge. It is known, of course, that
there exist different approaches for gauge covariant formulations of
non-abelian gauge theories.~\cite{gaugeonYM,YTM,gaugeonBRS2} \
However, they are not suitable to study the linear covariant gauge.
Therefore we do not go into the details of these approaches.  The
final section contains a summary.  In the appendix, we present an
explanation of how our method can also be understood from the point of
view of the non-abelian generalization of the model considered by
Kashiwa in Ref.~\citen{brstgauss}. Some results given there are found
to be useful in the understanding we describe in the main text, in
particular the BRST invariance of the extended system.

\section{'t~Hooft average as a path integral over gauge degrees of
  freedom}
\label{sec:thooft}

The Faddeev-Popov path integral\footnote{ As for the use of continual
  representation of PI's in this paper, it should be understood that
  we define them by their Euclidean version with the initial and final
  times, $t_{i}$ and $t_{f}$, and then taking the limits
  $t_{i}\to-\infty$ and $t_{f}\to+\infty$ in the end.}~(FPPI) in the
Landau gauge is given by
\begin{equation}
  \label{eq:thooft01}
  Z=\int\!\!\!{\mathcal D}A\,
  \delta(\partial^{\mu}A_{\mu})\varDelta[A]
  e^{iS[A]},
\end{equation}
where $S[A]$ is the gauge invariant action\footnote{ We employ matrix
  notation for quantities that take values in a group and its Lie
  algebra without expressing the group indices explicitly.} of the
(non-)abelian gauge field and $\varDelta[A]$ represents the explicit
form of the FP determinant $\varDelta_{\mathrm{FP}}[A,f]$ ($f=0$ in
the present case), defined by
\begin{equation}
  \label{eq:fpdet00}
  1=
  \varDelta_{\mathrm{FP}}[A,f]\int\!\!{\mathcal
    D}\theta\,\delta(\partial^{\mu}A_{\mu}^{\theta}-f),\quad
  A_{\mu}^{\theta}=\theta A_{\mu}\theta^{-1}+i\theta\partial_{\mu}\theta^{-1}.
\end{equation}
Because of the existence of $\delta(\partial^{\mu}A_{\mu})$ in
Eq.~\eqref{eq:thooft01}, it is evaluated to be
$\varDelta[A]=\left\vert\det(\partial^{\mu}D_{\mu})\right\vert$.  For
the reason explained below, we assume periodic boundary condition
(PBC) for all variables in the PI given in Eq.~\eqref{eq:thooft01}.

The above FPPI of the Landau gauge can be transformed to that for
other covariant gauges by making use of the 't~Hooft averaging: (i)
replacing the constraint $\delta(\partial^{\mu}A_{\mu})$ by
$\delta(\partial^{\mu}A_{\mu}-f)$ with an arbitrary $c$-number
function $f$, and (ii) inserting the Gaussian identity by regarding
$f$ introduced in the first step as identical to the integration
variable in
\begin{equation}
  \label{eq:gaussian}
  1=\int\!\!{\mathcal D}f\,\exp\left(-\frac{i}{2\alpha}
    \int\!\!d^{4}x\,f(x)^{2}\right).
\end{equation}
Then, the FPPI in Eq. \eqref{eq:thooft01} is rewritten as that
suitable for the $\alpha$-gauge,
\begin{equation}
  \label{eq:thooft02}
  Z=\int\!\!\!{\mathcal D}A\, \varDelta[A]
  e^{iS[A]-i\int\!d^{4}x\,(\partial^{\mu}A_{\mu})^{2}/(2\alpha)}.
\end{equation}

To elucidate the meaning of 't~Hooft's technique, let us suppose that
the $c$-number function $f$ in the constraint is related to a gauge
transformation through
$\partial^{\mu}A_{\mu}-f=\partial^{\mu}A_{\mu}^{g}$, where
$A_{\mu}^{g}$ is given by
\begin{gather}
    A_{\mu}^{g}=gA_{\mu}g^{-1}+ig\partial_{\mu}g^{-1}=
    A_{\mu}+igD_{\mu}g^{-1},\nonumber\\
    D_{\mu}g^{-1}=\partial_{\mu}g^{-1}-i[A_{\mu},g^{-1}].
  \label{eq:thooft03}
\end{gather}
We thus set $f=-i\partial^{\mu}(gD_{\mu}g^{-1})$ in the procedure of
the 't~Hooft average above.

Because we fix the gauge field $A_{\mu}$ by the variation $\delta
f=-i\partial^{\mu}(D_{\mu}^{g}(g\delta g^{-1}))$, in which
$D_{\mu}^{g}(g\delta g^{-1})=\partial_{\mu}(g\delta
g^{-1})-i[A_{\mu}^{g},(g\delta g^{-1})]$, and the invariant measure on
the group manifold is obtained from the volume form, which is defined
by the wedge product that consists of $gdg^{-1}$, we can rewrite the
Gaussian identity utilized above as
\begin{equation}
  \label{eq:gaussiang}
  1=\int\!\!{\mathcal D}g\,
  \varDelta[A^{g}]\,
  \exp\left(-\frac{i}{2\alpha}
    \int\!\!d^{4}x\,\{-i\partial^{\mu}(gD_{\mu}g^{-1})\}^{2}\right),
\end{equation}
where
$\varDelta[A^{g}]=\left\vert\det(\partial^{\mu}D_{\mu}^{g})\right\vert$.
We therefore find
\begin{align}
    Z=&\int\!\!{\mathcal D}A\,{\mathcal D}B\,{\mathcal D}g\,
    \varDelta[A]\,\varDelta[A^{g}]
    \exp\left({iS[A]}+i\int\!\!d^{4}x\,
      B\partial^{\mu}A_{\mu}^{g}\right)\nonumber\\
    &\times \exp\left(-\frac{i}{2\alpha}\int\!\!d^{4}x\,
      \left\{-i\partial^{\mu}(gD_{\mu}g^{-1})\right\}^{2}\right),
  \label{eq:thooft04}
\end{align}
which, of course (if the integrations with respect to $g$ are carried
out first), results in a PI for the $\alpha$-gauge with the
Nakanishi-Lautrup~(NL)~\cite{NL} field $B$. Although the PI given in
Eq.~\eqref{eq:thooft04}, obtained by setting
$f=-i\partial^{\mu}(gD_{\mu}g^{-1})$ in Eq.~\eqref{eq:gaussian}, is
merely a re-expression of 't~Hooft's original prescription, as we see
below, this expression makes it clear that the 't~Hooft average can be
understood as a PI over gauge degrees of freedom.

Here we comment on the boundary conditions of the PI given in
Eq.~\eqref{eq:thooft01}. In the original formulation of 't~Hooft's
technique, the identity of the functional Gaussian integral does not
require any boundary conditions, because it can be defined as a
product of functional integrals over variables on a surface of
constant $x_{0}$,
\begin{equation}
  \label{eq:thooft006}
  1=\int\!\!\prod_{\boldsymbol{x}\in\boldsymbol{R}^{3}}
  \left[\sqrt{\frac{i\Delta x_{0}}{2\pi\alpha}}\,
    df(\boldsymbol{x},x_{0})\right]
  \,\exp\left(-\frac{i\Delta x_{0}}{2\alpha}
    \int\!\!d^{3}x\,f(x)^{2}\right),
\end{equation}
where $\Delta x_{0}$ is the infinitesimal increment of $x_{0}$ in the
discretized PI. If the number of equal-time surfaces in such a PI is
$n+1$, specified by $t_{i}=x_{0}(0)$, $x_{0}(1),\,\dots,\,x_{0}(n-1)$,
and $x_{0}(n)=t_{f}$, in this order, we have $n$ integrals with
respect to $f(\boldsymbol{x},x_{0}(j))$~($j=1,\,2,\,\dots,\,n$),
corresponding to $n$ delta functions.  Then, setting
$f=-i\partial^{\mu}(gD_{\mu}g^{-1})$, the initial and final values of
$g$ enter the integrand, because $f$ includes a time derivative. They
must be connected by some condition, as otherwise the numbers of the
variables $f$ and $g$ do not match.  For a PI with PBC, we can simply
set the two boundary values equal. In this way, we can change the
gauge conditions in a relatively simple way in the trace formulae. The
reason for this simplicity is that the value of a trace is merely a
number, and therefore we can add any expression to it, provided that
the value of this expression vanishes. The cancellation
mechanism~\cite{KO} resulting from the BRST invariance~\cite{BRST}
ensures that contributions from unphysical degrees of freedom to a PI
with PBC satisfy this condition. In contrast to this simple case, the
situation becomes much more complicated for PIs with other boundary
conditions. Dealing with such a delicate issue is not the aim of this
paper. We thus restrict our inquiry here to the case of PIs with PBC.
Note, however, that naive use of the PBC may introduce complications
due to the existence of zero modes in cases of massless fields. Hence,
we must treat them carefully. (See Ref.~\citen{KS}, for example, for
details regarding this point.)

The Gaussian identity \eqref{eq:gaussiang} can be regarded as a
non-abelian generalization of the BRST invariant version of the
Froissart model~\cite{froissart} studied by Kashiwa in
Ref.~\citen{brstgauss} in detail. The map given by
$f=-i\partial^{\mu}(gD_{\mu}g^{-1})$ is then considered to be a
Nicolai map.~\cite{nicolai} \ (See Appendix \ref{sec:appendixa} of
this paper and discussions in Ref.~\citen{brstgauss}.)

\section{Extended gauge symmetry}
\subsection{The system with the extended gauge symmetry in 't~Hooft's
  path integral}
\label{sec:newpi}

Although the equivalence of the expression in Eq.~\eqref{eq:thooft04}
with that in Eq.~\eqref{eq:thooft02} is evident, our method of
obtaining Eq.~\eqref{eq:thooft04} may seem heuristic. In particular,
the FP determinant $\varDelta[A,f]$ in Eq.~\eqref{eq:fpdet00} must be
equal to $\varDelta[A^{g}]$ if we obtain the constraint
$\delta(\partial^{\mu}A_{\mu}-f)=\delta(\partial^{\mu}A_{\mu}^{g})$
through a gauge transformation from Eq.~\eqref{eq:thooft01}. However,
we have set $f=-i\partial^{\mu}(gD_{\mu}g^{-1})$, simply leaving the
determinant $\varDelta[A]$ [which is obtained from
$\varDelta[A,f]\delta(\partial^{\mu}A_{\mu}-f)=\varDelta[A]\delta(\partial^{\mu}A_{\mu}-f)$
if $f$ is a $c$-number function] unchanged in the derivation of the PI
of Eq.~\eqref{eq:thooft04}. Therefore we need to verify the derivation
of Eq.~\eqref{eq:thooft04} above. For this purpose, below we present
another derivation of this PI. However, before doing so, it is useful
to clarify the advantageous features of the PI in
Eq.~\eqref{eq:thooft04}.  For this purpose, here we examine the
$\alpha$ dependence of this PI.  If we wish to obtain a PI, as
explained above, for the $\alpha$-gauge, we first complete the square
of $f+\alpha B$ before carrying out the PI of $g$ with the measure
${\mathcal D}f={\mathcal D}g\,\varDelta[A^{g}]$.  Alternatively,
we may first perform the gauge transformation $A_{\mu}^{g}\mapsto
A_{\mu}$, under which we have $f\mapsto
i\partial^{\mu}(g^{-1}D_{\mu}g)$. We then find
\begin{align}
    Z=&\int\!\!{\mathcal D}A\,{\mathcal D}B\,{\mathcal D}g\,
    \varDelta[A^{g^{-1}}]\,\varDelta[A]
    \exp\left({iS[A]}+i\int\!\!d^{4}x\,B\partial^{\mu}A_{\mu}\right)\nonumber\\
    &\times \exp\left(-\frac{i}{2\alpha}\int\!\!d^{4}x\,
      \left\{i\partial^{\mu}(g^{-1}D_{\mu}g)\right\}^{2}\right).
  \label{eq:landau-gauge}
\end{align}
Then computing the PI of $g$, we find that the quantity in
Eq.~\eqref{eq:landau-gauge} is equal to the original FPPI of
Eq.~\eqref{eq:thooft01}. In other words, we can consider the 't~Hooft
average as the insertion of the identity
\begin{equation}
  \label{eq:gaussiang01}
  1=\int\!\!{\mathcal D}g\,
  \varDelta[A^{g^{-1}}]\,
  \exp\left(-\frac{i}{2\alpha}
    \int\!\!d^{4}x\,\{i\partial^{\mu}(g^{-1}D_{\mu}g)\}^{2}\right),
\end{equation}
which is, of course, equivalent to Eq.~\eqref{eq:gaussiang}, into the
original FPPI Eq.~\eqref{eq:thooft01} for the Landau gauge.  We have
thus confirmed that $Z$ is actually independent of $\alpha$.  It is
interesting that the PI in Eq.~\eqref{eq:thooft04} simultaneously
represents PIs for both the Landau gauge and the $\alpha$-gauge,
provided that we first integrate over the additional degrees of
freedom. Because $\alpha$ is arbitrary, we expect that the PI given by
Eq.~\eqref{eq:thooft04} with extended degrees of freedom contains all
linear covariant gauges, given by $\partial^{\mu}A_{\mu}+\alpha B=0$,
for the original variables.

We have performed the gauge transformation of $A_{\mu}$ above in order
to return to the PI for the Landau gauge from Eq.~\eqref{eq:thooft04}.
If we extend the gauge transformation to include that for $g$ in
addition to $A_{\mu}\mapsto A_{\mu}^{h}$, in order for $A_{\mu}^{g}$
to be invariant, we observe that the action in the first line of
Eq.~\eqref{eq:thooft04}, except for the gauge-variant factor
$\varDelta[A]$, possesses gauge invariance under the extended
transformations~\cite{exgauge}
\begin{equation}
  \label{eq:exgauge}
  A\mapsto  A^{h},\quad
  g\mapsto  gh^{-1},
\end{equation}
where $h$ takes values in the gauge group. Then, the second line of
Eq.~\eqref{eq:thooft04} combined with $\varDelta[A]$ is recognized as
a gauge fixing for this gauge invariance. We thus realize that the
procedure of the 't~Hooft averaging consists essentially of the
following two steps: (i) an extension of the degrees of freedom that
compensates for the gauge degrees of the gauge field through the above
gauge invariant implementation and (ii) a gauge fixing for this
extended gauge symmetry.

We now attempt to extend the observation above to the formulation of
the PI in Eq.~\eqref{eq:thooft04}. We first rewrite
Eq.~\eqref{eq:thooft01} by replacing $A_{\mu}$ with $A_{\mu}^{g}$ and
define the formally divergent PI
\begin{equation}
  \label{eq:divpi01}
  Z_{\mathrm{div}}=
  \int\!\!
  {\mathcal D}A\,{\mathcal D}B\,{\mathcal D}g\,
  I[A,B,g]
  \left(=\int\!\!{\mathcal D}g\,\int\!\!
    {\mathcal D}A\,{\mathcal D}B\,
    I[A,B,1]\right),
\end{equation}
in which the gauge invariant functional $I[A,B,g]$ is given by
\begin{equation}
  \label{eq:divpi02}
  I[A,B,g]=\varDelta[A^{g}]
  \exp\left(iS[A]+i\int\!\!d^{4}x\,
    B\partial^{\mu}A_{\mu}^{g}\right).
\end{equation}

The FP trick for this PI with gauge invariance is implemented by
inserting the identity
\begin{equation}
  \label{eq:fptrick01}
  1=\varDelta_{\mathrm{FP}}[A,g,C]\int\!\!{\mathcal D}h\,
  \delta(f[gh,A^{h^{-1}}]-C),
\end{equation}
where $C$ is an arbitrary $c$-number function, and we have written
$f=-i\partial^{\mu}(gD_{\mu}g^{-1})$ as $f[g,A]$, taking its
functional dependence into account.  By making use of the invariance
of $I[A,B,g]$ and $\varDelta_{\mathrm{FP}}[A,g,C]$ under
Eq.~\eqref{eq:exgauge}, we can factorize $\int\!\!{\mathcal
  D}h(=\infty)$ as usual to obtain
\begin{equation}
  \label{eq:newpi01}
  Z=\int\!\!
  {\mathcal D}A\,{\mathcal D}B\,{\mathcal D}g\,
  I[A,B,g]\varDelta_{\mathrm{FP}}[A,g,C]\delta(f[g,A]-C).
\end{equation}
Then, because we have
\begin{equation}
  \label{eq:newfpdet01}
  \varDelta_{\mathrm{FP}}[A,g,C]\delta(f[g,A]-C)=\varDelta[A]\delta(f[g,A]-C)
\end{equation}
and Eq.~\eqref{eq:newpi01} is independent of $C$, we may insert the
Gaussian identity for $C$. We then find that this PI results in
Eq.~\eqref{eq:thooft04}. We have thus completed the explanation of
Eq.~\eqref{eq:thooft04} from our new point of view.

Extending the degrees of freedom for the extended gauge symmetry, we
introduce new fields, which require an extension of the Hilbert space
from that of the original degrees of freedom. To examine the structure
of this extended Hilbert space, let us Fourier transform the second
line of Eq.~\eqref{eq:thooft04}. This gives
\begin{align}
    Z=&\int\!\!{\mathcal D}A\,{\mathcal D}B\, {\mathcal D}g\,{\mathcal
      D}\varPhi\, \varDelta[A]\,\varDelta[A^{g}]
    \exp\left(iS[A]+i\int\!\!d^{4}x\,
      B\partial^{\mu}A_{\mu}^{g}\right)\nonumber\\
    &\times \exp\left(i
      \int\!\!d^{4}x\,\left\{-i\varPhi\partial^{\mu}(gD_{\mu}g^{-1})+
        \frac{\alpha}{2}\varPhi^{2}\right\}\right).
  \label{eq:thooft05}
\end{align}
Then, if we set $g=e^{i\varTheta}$ and shift the field $\varPhi$ to
$\varPhi+B$, we find that the quadratic part of the Lagrangian is
given by
\begin{equation}
  \label{eq:thooft06}
  {\mathcal L}_{\mathrm{G}}^{(0)}=
  -\varPhi\partial^{\mu}\partial_{\mu}\varTheta
  +\frac{\alpha}{2}(\varPhi+B)^{2}
\end{equation}
for these new variables. Because both $\varPhi$ and $B$ are subject to
the d'Alembert equation at the tree level, $\varTheta$ is regarded as
a dipole ghost, except in the case $\alpha=0$.  Hence we need an
indefinite metric for the sector of $\varTheta$ and $\varPhi$. This
unphysical sector is accompanied by that for ghost fermions coming
from\footnote{We assume that $\varDelta[A]$ and $\varDelta[A^{g}]$ do
  not vanish, so that sign changes do not occur. In this case, the
  Gribov problem~\cite{Gribov} can be avoided in the perturbative
  definition of these quantities.}
\begin{equation}
  \label{eq:thooft07}
  \varDelta [A]=\int\!\!{\mathcal D}\bar{c}\,{\mathcal D}c\,
  \exp\left(i
    \int\!\!d^{4}x\,i\bar{c}\partial^{\mu}D_{\mu}c\right),
\end{equation}
and also from
\begin{equation}
  \label{eq:thooft007}
  \varDelta [A^{g}]=\int\!\!{\mathcal D}\bar{\eta}\,{\mathcal D}\eta\,
  \exp\left(i
    \int\!\!d^{4}x\,i\bar{\eta}\partial^{\mu}D_{\mu}^{g}\eta^{g}\right),
\end{equation}
where $\eta^{g}=g\eta g^{-1}$, as well. Combining these with the other
factors in Eq.~\eqref{eq:thooft05}, we obtain the Lagrangian
\begin{equation}
  \label{eq:thooft08}
  {\mathcal L}=
  -\frac{1}{4}F^{\mu\,\nu}F_{\mu\,\nu} +B\partial^{\mu}A_{\mu}^{g}
  +i\bar{\eta}\partial^{\mu}D_{\mu}^{g}\eta^{g}+
  \varPhi f[g,A]+
  \frac{\alpha}{2}\varPhi^{2}+
  i\bar{c}\partial^{\mu}D_{\mu}c,
\end{equation}
where
$F_{\mu\,\nu}=\partial_{\mu}A_{\nu}-\partial_{\nu}A_{\mu}-i[A_{\mu},\,A_{\nu}]$.

If we arrange this Lagrangian as
\begin{align}
    {\mathcal
      L}=&-\frac{1}{4}F^{\mu\,\nu}F_{\mu\,\nu}+B\partial^{\mu}A_{\mu}
    +\frac{\alpha}{2}B^{2}+
    i\bar{c}\partial^{\mu}D_{\mu}c\nonumber\\
    &+i\bar{\eta}\partial^{\mu}D_{\mu}^{g}\eta^{g}+
    \frac{\alpha}{2}\left(\varPhi+\frac{1}{\alpha}f[g,A]\right)^{2}-
    \frac{1}{2\alpha}(f[g,A]+\alpha B)^{2},
  \label{eq:lagalpha}
\end{align}
we see that the integration over $g$ and $\varPhi$ in combination with
that over $\bar{\eta}$ and $\eta$ yields the Lagrangian in the
$\alpha$-gauge for the original variables. On the other hand, we may
rewrite the Lagrangian, first performing the gauge transform of
$A_{\mu}$ by taking $A_{\mu}\mapsto A_{\mu}^{g^{-1}}$ and
$(\bar{\eta},\eta^{g})\mapsto(\bar{c},c)$, together with
$(\bar{c},c)\mapsto(\bar{\eta},\eta^{g^{-1}})$. This yields
\begin{align}
    {\mathcal L}\mapsto{\mathcal
      L}{'}=&-\frac{1}{4}F^{\mu\,\nu}F_{\mu\,\nu}+B\partial^{\mu}A_{\mu}
    +i\bar{c}\partial^{\mu}D_{\mu}c\nonumber\\
    &+i\bar{\eta}\partial^{\mu}D_{\mu}^{g^{-1}}\eta^{g^{-1}}-\varPhi
    f[g^{-1},A]+ \frac{\alpha}{2}\varPhi^{2},
  \label{eq:laglandau}
\end{align}
corresponding to the method of obtaining Eq.~\eqref{eq:thooft01} from
Eq.~\eqref{eq:thooft04} demonstrated above. The important point here
is that the PI given by Eq.~\eqref{eq:thooft04} simultaneously
contains these two systems, described by Lagrangians
Eq.~\eqref{eq:lagalpha} and Eq.~\eqref{eq:laglandau}. It is noteworthy
that such a structure, i.e. a hybrid of the Landau gauge and the
$\alpha$-gauge, of the PI in Eq.~\eqref{eq:thooft04} can be
constructed only after our identification of the unknown $c$-number
function $f$ in the 't~Hooft average with
$f[g,A]=-i\partial^{\mu}(gD_{\mu}g^{-1})$, which allows us to
interpret the system being equipped with the extended gauge
transformation \eqref{eq:exgauge}.

\subsection{BRST invariance and the structure of the total Hilbert
  space}
\label{sec:brst}
As we saw in the preceding section, the 't~Hooft average can be
interpreted as a PI with gauge fixing for the system with extended
gauge invariance. Since we have formulated this PI by means of the FP
trick, as usual, it is natural to conjecture the BRST invariance for
this gauge fixing procedure. We show in this section that this is
indeed the case, and, in addition to the usual BRST symmetry, there
exists another BRST symmetry for the system described by the
Lagrangian \eqref{eq:thooft08}.

By replacing the $c$-number function $\theta(x)$ in the gauge
transformations $A_{\mu}\mapsto A_{\mu}+D_{\mu}\theta$ and $g\mapsto
g-ig\theta$ with $\lambda c(x)$, we observe that the Lagrangian in
Eq.~\eqref{eq:thooft08} is invariant under the BRST transformation
\begin{gather}
    \delta A_{\mu}=\lambda D_{\mu}c,\quad
    \delta\bar{c}=i\lambda\varPhi,\quad -ig\delta g^{-1}=\lambda
    gcg^{-1},\quad
    \delta c=i\lambda c^{2},\nonumber\\
    \delta\eta=i\lambda\{c,\eta\},\quad \delta\bar{\eta} =\delta
    B=\delta\varPhi=0,
  \label{eq:thooft09}
  \end{gather}
where $\lambda$ is a Grassmann parameter. Apparently, the BRST
invariance mentioned above corresponds to the gauge fixing of the extended
gauge symmetry.  In addition to this usual BRST invariance, there
exists the transformation
\begin{gather}
    -ig\tilde{\delta} g^{-1}=\tilde{\lambda}g\eta g^{-1}
    =\tilde{\lambda}\eta^{g},\quad
    \tilde{\delta}\bar{\eta}=i\tilde{\lambda}(\varPhi-B),\quad
    \tilde{\delta}\eta=i\tilde{\lambda}\eta^{2},\nonumber\\
    \tilde{\delta} A_{\mu}=\tilde{\delta}\bar{c}= \tilde{\delta} c=
    \tilde{\delta} B=\tilde{\delta}\varPhi=0,
  \label{eq:thooft009}
\end{gather}
under which the system remains invariant.  This additional BRST
invariance originates from that given in the second line of
Eq.~\eqref{eq:laglandau}, and it is a consequence of the trivial
nature of the Gaussian identity \eqref{eq:gaussiang} [or
\eqref{eq:gaussiang01}] when expressed as a PI with ghost fermions.
(See Appendix \ref{sec:appendixa} for details.)

Setting $\delta=\lambda\boldsymbol{\delta}$ and
$\tilde{\delta}=\tilde{\lambda}\tilde{\boldsymbol{\delta}}$ in
Eqs.~\eqref{eq:thooft09} and \eqref{eq:thooft009}, respectively, we
can write the Lagrangian as
\begin{equation}
  \label{eq:brstlag01}
  {\mathcal L}=
  -\frac{1}{4}F^{\mu\,\nu}F_{\mu\,\nu} +B\partial^{\mu}A_{\mu}^{g}
  +i\bar{\eta}\partial^{\mu}D_{\mu}^{g}\eta^{g}
  -i\boldsymbol{\delta}\left[
    \bar{c}\left(f[g,A]+\frac{\alpha}{2}\varPhi\right)\right]
\end{equation}
or as
\begin{equation}
  \label{eq:brstlag02}
  {\mathcal L}=
  -\frac{1}{4}F^{\mu\,\nu}F_{\mu\,\nu} +B\partial^{\mu}A_{\mu}
  +i\bar{c}\partial^{\mu}D_{\mu}c+\frac{\alpha}{2}B^{2}
  -i\tilde{\boldsymbol{\delta}}\left[
    \bar{\eta}\left(f[g,A]+\frac{\alpha}{2}(\varPhi+B)\right)\right].
\end{equation}
Furthermore, if we write
$\boldsymbol{\delta}_{\mathrm{B}}=\boldsymbol{\delta}+\tilde{\boldsymbol{\delta}}$,
the Lagrangian can be expressed as
\begin{equation}
  \label{eq:brstlag03}
  {\mathcal L}=
  -\frac{1}{4}F^{\mu\,\nu}F_{\mu\,\nu}
  -i\boldsymbol{\delta}_{\mathrm{B}}\left[
    \bar{c}\left(\partial^{\mu}A_{\mu}+\frac{\alpha}{2}\varPhi\right)-
    \bar{\eta}\partial^{\mu}A_{\mu}^{g}\right].
\end{equation}
Note that we have defined $\boldsymbol{\delta}$ and
$\tilde{\boldsymbol{\delta}}$ to be nilpotent and also to anti-commute
with each other. Therefore $\boldsymbol{\delta}_{\mathrm{B}}$ is also
nilpotent. Next, we can define the BRST charges $Q$, $\tilde{Q}$ and
$Q_{\mathrm{B}}$ corresponding to $\boldsymbol{\delta}$,
$\tilde{\boldsymbol{\delta}}$ and $\boldsymbol{\delta}_{\mathrm{B}}$,
respectively. Then the state vectors destroyed by multiplying these
charges will be \textit{physical states}. Here, the meaning of {\em
  physical} needs to be explained. In view of
Eq.~\eqref{eq:brstlag03}, it is evident that the subspace specified by
the condition $Q_{\mathrm{B}}\vert\textrm{phys}\rangle=0$ is
equivalent to that defined by the physical state condition proposed by
Kugo and Ojima~\cite{KO} for the original degrees. This is the meaning
of the term \textit{physical} in application to $Q_{\mathrm{B}}$.
Then the expression of the Lagrangian given by
Eq.~\eqref{eq:brstlag02} reveals that there exists a local
decomposition of the total Hilbert space into subspaces, that
specified by $\tilde{Q}\vert\textrm{phys};\alpha\rangle=0$ and the
rest. Because the PI of the Lagrangian ${\mathcal L}_{\mathrm{ex}}$
given below with respect to $g$, $\varPhi$, $\eta$ and $\bar{\eta}$ is
trivial owing to the quartet mechanism, only the vacuum of these
extended degrees of freedom can be a positive normed and physical
state with respect to $\tilde{Q}$ in the subspace that describes the
system defined by
\begin{equation}
  \label{eq:brstlag04}
  {\mathcal L}_{\mathrm{ex}}=
  -i\tilde{\boldsymbol{\delta}}\left[
    \bar{\eta}\left(f[g,A]+\frac{\alpha}{2}(\varPhi+B)\right)\right]
\end{equation}
for a fixed configuration (i.e. the $\alpha$-gauge in the present
case) of $A_{\mu}$ and $B$. Therefore, $\tilde{Q}$ defines the Hilbert
space for the $\alpha$-gauge of the original degrees of freedom as its
invariant subspace. For this reason, we have written the condition for
$\tilde{Q}$ as $\tilde{Q}\vert\textrm{phys};\alpha\rangle=0$ above.
Then, integrating out the extended degrees of freedom, we obtain the
reduced system for the original variables in the $\alpha$-gauge.
Because, in the course of this reduction, the BRST charge $Q$ is
reduced to $Q_{\mathrm{KO}}$ (the BRST charge of Kugo and Ojima), we
observe that the total BRST charge $Q_{\mathrm{B}}=Q+\tilde{Q}$ is the
proper extension of $Q_{\mathrm{KO}}$ needed to fit the extended gauge
symmetry.  We note here that the structure of the total Hilbert space
for the extended system is quite similar to that found by Koseki, Sato
and Endo in Ref.~\citen{gaugeonKSE} for the BRST invariant version of
Yokoyama's gaugeon formalism. This similarity is discussed in more
detail in \S~\ref{sec:gaugeon}.

In the same way, $Q$ defines its own invariant subspace, according to
the decomposition of the Lagrangian given by Eq.~\eqref{eq:brstlag01}.
However, it is impossible to carry out the integrations with respect
to the additional degrees of freedom in this form because the subspace
is the Hilbert space of $A_{\mu}^{g}$ and $B$ (along with their
ghosts) for the Landau gauge. We therefore need to perform the same
transformations as in the case that we obtained
Eq.~\eqref{eq:laglandau} in order to separate the additional degrees
of freedom from the original ones. Then the corresponding
decomposition of the Lagrangian becomes
\begin{equation}
  \label{eq:brstlag05}
  {\mathcal L}{'}=
  -\frac{1}{4}F^{\mu\,\nu}F_{\mu\,\nu} +B\partial^{\mu}A_{\mu}
  +i\bar{c}\partial^{\mu}D_{\mu}c
  -i\tilde{\boldsymbol{\delta}}{'}\left[
    \bar{\eta}\left(-f[g^{-1},A]+\frac{\alpha}{2}\varPhi\right)\right]
\end{equation}
in which $\tilde{\boldsymbol{\delta}}{'}$ differs from
$\tilde{\boldsymbol{\delta}}$ due to the change of the rule for
$\bar{\eta}$ in Eq.~\eqref{eq:thooft009} to
$\tilde{\boldsymbol{\delta}}{'}\bar{\eta}=i\varPhi$, but is, of
course, nilpotent and anti-commutes with
$\boldsymbol{\delta}_{\mathrm{B}}$. Note that the positions of the two
pairs of ghost fermions, $(\bar{c},c)$ and $(\bar{\eta},\eta)$, are
exchanged under the change of variables that brings ${\mathcal L}$ to
${\mathcal L}{'}$.  With this change of variables, the BRST charge $Q$
is transformed to $\tilde{Q}{'}$, corresponding to
$\tilde{\boldsymbol{\delta}}{'}$ above.  The transformed charge
$\tilde{Q}{'}$ then defines the Hilbert space for the Landau gauge of
the original variables as its invariant subspace.  As the counterpart
to the transformation $Q\mapsto\tilde{Q}{'}$, $\tilde{Q}$ of the
original system transforms to $Q{'}$, which is such that the relation
$Q_{\mathrm{B}}=Q+\tilde{Q}=Q{'}+\tilde{Q}{'}$ holds. Hence we again
observe the hybrid of the Landau gauge and the $\alpha$-gauge of the
original degrees of freedom in the structure of the total Hilbert
space.

\subsection{Gauge fixing without averaging over the $c$-number function}
\label{sec:delta}
To this point, we have presented our understanding of the 't~Hooft
average from the point of view of extended gauge symmetry and gauge
fixing for it. In the derivation of the PI given in
Eq.~\eqref{eq:thooft04}, however, we have used the same technique
(averaging with respect to a $c$-number function) again in
Eq.~\eqref{eq:newpi01} to find the Gaussian weight in the second line
of Eq.~\eqref{eq:thooft04}. In this sense, we have not yet realized
our entire goal for this paper. Here we show that we can avoid the use
of this technique and discuss the difference between this new
prescription and that utilized in previous sections.

Returning to the divergent PI \eqref{eq:divpi01} considered in
\S~\ref{sec:newpi}, let us reconsider the use of the FP trick for
$Z_{\mathrm{div}}$. By making use of the facts that (i) the identity
\eqref{eq:fptrick01} holds for any $c$-number function $C$ and (ii)
the FP determinant $\varDelta_{\mathrm{FP}}[A,g,C]$ becomes
$\varDelta[A]$ in front of $\delta(f[g,A]-C)$, we have multiplied
\eqref{eq:newpi01} by a Gaussian identity of $C$, regarding $C$ as
identical to the integration variable of the Gaussian identity. These
facts also ensure the validity of using
\begin{equation}
  \label{eq:fptrick02}
  1=\varDelta_{\mathrm{FP}}[A,g,-B/2]\int\!\!{\mathcal D}h\,
  \delta\left(f[gh,A^{h^{-1}}]+\frac{\alpha}{2}B\right)
\end{equation}
instead of Eq.~\eqref{eq:fptrick01} and the Gaussian averaging
with respect to $C$ afterward. Factorizing out the gauge volume from
$Z_{\mathrm{div}}$ again, we obtain
\begin{equation}
  \label{eq:typeII01}
  Z=\int\!\!
  {\mathcal D}A\,{\mathcal D}B\,{\mathcal D}g\,
  I[A,B,g]\varDelta[A]
  \delta\left(f[g,A]+\frac{\alpha}{2}B\right).
\end{equation}
Then, because by integrating $g$ out with the measure ${\mathcal
  D}f={\mathcal D}g\varDelta[A^{g}]$, we immediately return to
Eq.~\eqref{eq:thooft02} for the $\alpha$-gauge, we can easily check
that $Z$ given above is also equivalent to the PI for the original
degrees of freedom. We are therefore convinced that the PI given by
Eq.~\eqref{eq:typeII01}, with the extended degrees of freedom, is
useful for another gauge fixing procedure of the extended gauge
invariance, and we can avoid using the 't~Hooft average with this new
prescription of the gauge fixing.

Let us now consider the Lagrangian in the PI given in
Eq.~\eqref{eq:typeII01} from the point of view of the BRST invariance.
If we express the Fourier transform of the delta function in
Eq.~\eqref{eq:typeII01} as a functional integral with respect to
$\varPhi$, this Lagrangian reads
\begin{equation}
  \label{eq:typeIIlag01}
  {\mathcal L}_{\delta}=
  -\frac{1}{4}F^{\mu\,\nu}F_{\mu\,\nu} +B\partial^{\mu}A_{\mu}^{g}
  +i\bar{\eta}\partial^{\mu}D_{\mu}^{g}\eta^{g}+
  \varPhi f[g,A]+\frac{\alpha}{2}\varPhi B+
  i\bar{c}\partial^{\mu}D_{\mu}c.
\end{equation}
As will become evident, the difference between this Lagrangian and
that in Eq.~\eqref{eq:thooft08} appears only in the term proportional
to $\alpha$. Hence, we can define BRST transformations that are
identical to those given in Eqs.~\eqref{eq:thooft09} and
\eqref{eq:thooft009}.  Accordingly, we obtain similar decompositions of
this Lagrangian, given by
\begin{equation}
  \label{eq:typeIIlag02}
  {\mathcal L}_{\delta}=
  -\frac{1}{4}F^{\mu\,\nu}F_{\mu\,\nu} +B\partial^{\mu}A_{\mu}^{g}
  +i\bar{\eta}\partial^{\mu}D_{\mu}^{g}\eta^{g}
  -i\boldsymbol{\delta}\left[
    \bar{c}\left(f[g,A]+\frac{\alpha}{2}B\right)\right]
\end{equation}
and
\begin{equation}
  \label{eq:typeIIlag03}
  {\mathcal L}_{\delta}=
  -\frac{1}{4}F^{\mu\,\nu}F_{\mu\,\nu} +B\partial^{\mu}A_{\mu}
  +i\bar{c}\partial^{\mu}D_{\mu}c+\frac{\alpha}{2}B^{2}
  -i\tilde{\boldsymbol{\delta}}\left[
    \bar{\eta}\left(f[g,A]+\frac{\alpha}{2}B\right)\right],
\end{equation}
corresponding to Eqs.~\eqref{eq:brstlag01} and \eqref{eq:brstlag02},
respectively.  In the same way, in accordance with the nilpotency of
the total BRST transformation
$\boldsymbol{\delta}_{\mathrm{B}}=\boldsymbol{\delta}+\tilde{\boldsymbol{\delta}}$,
we can rewrite the Lagrangian as
\begin{equation}
  \label{eq:typeIIlag04}
  {\mathcal L}_{\delta}=
  -\frac{1}{4}F^{\mu\,\nu}F_{\mu\,\nu}
  -i\boldsymbol{\delta}_{\mathrm{B}}\left[
    \bar{c}\left(\partial^{\mu}A_{\mu}+\frac{\alpha}{2}B\right)-
    \bar{\eta}\partial^{\mu}A_{\mu}^{g}\right].
\end{equation}
In analogy to the change of the Lagrangian ${\mathcal L}$ to
${\mathcal L}{'}$, we can make a change of variables in ${\mathcal
  L}_{\delta}$ appearing in Eq.~\eqref{eq:typeIIlag02} to obtain
\begin{equation}
  \label{eq:typeIIlag05}
  {\mathcal L}_{\delta}{'}=
  -\frac{1}{4}F^{\mu\,\nu}F_{\mu\,\nu} +B\partial^{\mu}A_{\mu}
  +i\bar{c}\partial^{\mu}D_{\mu}c
  -i\tilde{\boldsymbol{\delta}}{'}\left[
    \bar{\eta}\left(-f[g^{-1},A]+\frac{\alpha}{2}B\right)\right],
\end{equation}
where $\tilde{\boldsymbol{\delta}}{'}$ is the same as that defined
above.  Therefore we observe that the total Hilbert space of the
extended system described by the Lagrangian \eqref{eq:typeIIlag01} has
exactly the same structure as that described by the Lagrangian
\eqref{eq:thooft09}.

We have thus obtained another method of gauge fixing for the extended
gauge symmetry. As explained at the beginning of this section, the
advantage of this second method of gauge fixing is that, with it, we
never need to carry out the averaging according to the Gaussian weight
with respect to an unknown $c$-number field.  Hence we have realized
the main goal of this paper by formulating this new method. It is
important, however, to note that the same prescription cannot be
applied to the case of the original form of the 't~Hooft average; it
can be applied only after the extension of the degrees of freedom
needed to obtain the extended gauge symmetry.  In this regard, it is
interesting that, as was pointed out by Nakanishi~\cite{NN} and also
shown by Yokoyama~\cite{gaugeon} for the NL formalism~\cite{NL} of
QED, we cannot change the gauge parameter $\alpha$ in a consistent way
without introducing additional degrees of freedom (a gaugeon and its
associated field) for the quantum gauge degree of freedom. Therefore,
our formulation of the PI with additional degrees of freedom may have
some connection to the gaugeon
formalism.~\cite{gaugeon,gaugeonIzawa,gaugeonKSE,gaugeonEndo} \ This
is the subject of the next section. However, before closing this
section, we give some discussion of the PIs in
Eqs.~\eqref{eq:thooft04} and \eqref{eq:typeII01}. Interestingly, the
resemblance of the decomposition of the Lagrangian in
Eq.~\eqref{eq:typeIIlag05} to that in Eq.~\eqref{eq:brstlag05}
suggests yet another way of deriving these PIs.  As we have already
shown for the case of Eq.~\eqref{eq:thooft04}, when we write the
Lagrangian of the PI in the form of Eq.~\eqref{eq:brstlag05}, the
corresponding PI can be regarded as a product of $Z$ in
Eq.~\eqref{eq:thooft01} and the identity
\begin{equation}
  \label{eq:typeIidentity}
  1=\int\!\!{\mathcal D}g\,{\mathcal D}\varPhi\,
  {\mathcal D}\bar{\eta}\,{\mathcal D}\eta\,
  \exp\left(
    \int\!\!d^{4}x\,\tilde{\boldsymbol{\delta}}{'}\left[
      \bar{\eta}\left(-f[g^{-1},A]+\frac{\alpha}{2}\varPhi\right)\right]\right),
\end{equation}
provided that we perform the gauge transformation $A_{\mu}\mapsto
A_{\mu}^{g}$ afterward. In the same sense, the identity
\begin{equation}
  \label{eq:typeIIidentity}
  1=\int\!\!{\mathcal D}g\,{\mathcal D}\varPhi\,
  {\mathcal D}\bar{\eta}\,{\mathcal D}\eta\,
  \exp\left(
    \int\!\!d^{4}x\,\tilde{\boldsymbol{\delta}}{'}\left[
      \bar{\eta}\left(-f[g^{-1},A]+\frac{\alpha}{2}B\right)\right]\right)
\end{equation}
can be multiplied by $Z$ of Eq.~\eqref{eq:thooft01} to obtain the PI
in Eq.~\eqref{eq:typeII01}. We are, therefore, convinced that the
procedure of implementing the extended gauge symmetry and the gauge
fixing by means of the FP trick is equivalent to the multiplication of
an identity that is given by the product of the bosonic and fermionic
functional determinant from a PI of the extended degrees of freedom.
With this understanding, we can avoid the procedure of averaging over
an unknown $c$-number function, even for the case of the PI given in
Eq.~\eqref{eq:thooft04}.  If we start our formulation with PI in
Eq.~\eqref{eq:thooft02} for the $\alpha$-gauge and multiply
Eq.~\eqref{eq:typeIidentity} or \eqref{eq:typeIIidentity} after
replacing $\alpha$ in these identities with
$\delta\alpha=\alpha{'}-\alpha$, we obtain the PI of the extended
system as a hybrid of the $\alpha$-gauge and $\alpha{'}$-gauge.  This
extended PI then reduces to that suitable for the $\alpha{'}$-gauge of
the original system by integrating out the additional degrees of
freedom after the gauge transformation $A_{\mu}\mapsto A_{\mu}^{g}$.

\section{Relation to the gaugeon formalism}
\label{sec:gaugeon}
We start this section with a brief review of the essence of Yokoyama's
gaugeon formalism~\cite{gaugeon} for QED.  The gauge fixing term
$B\partial^{\mu}A_{\mu}$ is extended to
$B\partial^{\mu}A_{\mu}-\varphi\partial^{\mu}\partial_{\mu}\theta$,\footnote{The
  corresponding notation for the fields $(\theta,\varphi,B)$ in the
  original paper by Yokoyama~\cite{gaugeon} is $(B,B_{2},B_{1})$.
  Koseki et al. adopt the notation $(Y,Y_{*},B)$ in
  Ref.~\citen{gaugeonKSE}.}  and then the extended Lagrangian should be
invariant under the $q$-number gauge transformation given by
\begin{equation}
  \label{eq:gaugeon01}
  A_{\mu}\mapsto A_{\mu}+a\partial_{\mu}\theta,\quad
  \varphi\mapsto\varphi+aB,
\end{equation}
leaving $B$ and $\theta$ intact. (This is a global symmetry with the
global parameter $a$. The resemblance to the BRST invariance should be
noted.) Because $\theta$ describes the quantum gauge degree of freedom
of $A_{\mu}$, it is called a gaugeon field and appears in the extended
Lagrangian with its partner field $\varphi$. They add a term to the
Lagrangian that breaks the symmetry above, and we obtain
\begin{equation}
  \label{eq:gaugeon02}
  {\mathcal L}_{\mathrm{I}}={\mathcal L}_{0}+
  B\partial^{\mu}A_{\mu}-\varphi\partial^{\mu}\partial_{\mu}\theta+
  \frac{\epsilon_{1}}{2}(\varphi+a_{0}B)^{2},
\end{equation}
where ${\mathcal L}_{0}$ represents the gauge invariant Lagrangian of
genuine QED, for type I gaugeon formalism. For the case of the type II
gaugeon theory, the Lagrangian is given by
\begin{equation}
  \label{eq:gaugeon03}
  {\mathcal L}_{\mathrm{II}}={\mathcal L}_{0}+
  B\partial^{\mu}A_{\mu}-\varphi\partial^{\mu}\partial_{\mu}\theta+
  \frac{\epsilon_{2}}{2}(\varphi+a_{0}B)B.
\end{equation}
If we integrate out $\theta$ with $\varphi$, the Lagrangians of both
systems reduce to that of the original degrees of freedom for the
$\alpha$-gauge; $\alpha=\epsilon_{1}a_{0}^{2}$ for a type I system and
$\alpha=\epsilon_{2}a_{0}$ for a type II system.  However, if we carry
out the $q$-number gauge transformation given by
Eq.~\eqref{eq:gaugeon01} first, the gauge parameter $\alpha$ of the
resulting reduced system becomes $\alpha{'}=\epsilon_{1}(a+a_{0})^{2}$
for the type I case and $\alpha{'}=\epsilon_{2}(a+a_{0})$ for the type
II case. Therefore we can change the gauge parameter for the reduced
system by performing the $q$-number gauge transformation before
integrating out the additional degrees of freedom. The difference
between the type I and type II formalisms is in their rules for
changing $\alpha$ to $\alpha{'}$ via Eq.~\eqref{eq:gaugeon01}. For the
case of a type I system, the rule is
$\alpha{'}=(1+a/a_{0})^{2}\alpha$, while for a type II system, it is
$\alpha{'}=(1+a/a_{0})\alpha$.

The BRST invariant formulation of the gaugeon formalisms above were
given by Izawa for the type II case~\cite{gaugeonIzawa} and by Koseki,
Sato and Endo~\cite{gaugeonKSE} for both types of gaugeon formalism.
The generalization of the type I theory discussed above, in which the
rule for changing the gauge parameter in order to admit any real
values for $\alpha$, was also formulated by Endo.~\cite{gaugeonEndo} \
Let us examine here whether we can generalize the derivation of the PI
carried out by Koseki et al. in Ref.~\citen{gaugeonKSE} to non-abelian
gauge theories. The key to their derivation seems to be multiplication
by unity expressed by the right-hand side of an identity of the form
$1=\det\Box^{-1}\cdot\det\Box$, which should further be rewritten as a
PI with BRST invariance.  The corresponding identity in our case is
given by Eq.~\eqref{eq:typeIidentity}.  Then, we can follow the
procedure employed in Ref.~\citen{gaugeonKSE} to obtain a non-abelian
generalization of the ${\mathcal L}_{\mathrm{YK}}$ that includes three
parameters.  For the case of an abelian theory, Koseki et al. observed
that gaugeon formalism can be reproduced by setting these three
parameters as $\alpha_{1}=\pm1=\varepsilon$, $\alpha_{2}=\varepsilon
a$, $\alpha_{3}=\varepsilon a^{2}$ for the type I formalism and as
$\alpha_{1}=0$, $\alpha_{2}=1/2$, $\alpha_{3}=a$ for the type II
theory, respectively.  The corresponding Lagrangians in our
formulation are those appearing in Eqs.~\eqref{eq:thooft08} and
\eqref{eq:typeIIlag01}. In these Lagrangians, however, we do not have
any free parameter that represents $a$ in the gaugeon formalism.
Rather, we have to set $a=1$ as well as $\alpha_{1}=\alpha$ for
Eq.~\eqref{eq:thooft08} and $\alpha_{2}=\alpha/2$ for
Eq.~\eqref{eq:typeIIlag01}. Thus, our results partially generalize
those of Ref.~\citen{gaugeonKSE}. The reason why the three parameters
in gaugeon formalism of Yokoyama and Kubo are not useful in
non-abelian cases is explained below, but it is important to keep in
mind that we can always perform the $q$-number gauge transformation
\eqref{eq:gaugeon01} to eliminate the parameters $a_{0}$ in
Eqs.~\eqref{eq:gaugeon02} and \eqref{eq:gaugeon03}. From the point of
view of form invariance in the gaugeon formalism, the three parameters
are fundamental. But as seen below, the concept of form invariance
cannot be regarded as a generic one for linear covariant gauges when
we extend the formalism to non-abelian systems.

If the global symmetry under the $q$-number gauge transformation
\eqref{eq:gaugeon01} observed above generalizes to non-abelian cases
as well, our formulation of the PIs given in Eqs.~\eqref{eq:thooft04}
and \eqref{eq:typeII01} can be regarded as gaugeon formalisms for
non-abelian gauge theories. However, this is not the case: Due to the
fact that we need to change $A_{\mu}$ in $gD_{\mu}g^{-1}$, the
possible form given by
$B\partial^{\mu}A_{\mu}+i\varPhi\partial^{\mu}(gD_{\mu}g^{-1})$ for
the generalization of
$B\partial^{\mu}A_{\mu}-\varphi\partial^{\mu}\partial_{\mu}\theta$
cannot be invariant under any finite gauge transformation of $A_{\mu}$
in combination with a shift in $\varPhi$ proportional to $B$. To
resolve this problem, there can exist only one possibility, that is,
that we stipulate $g=e^{i\varTheta}$ to be infinitesimal and consider
the $q$-number gauge transformation within this infinitesimal
one-dimensional subgroup,
\begin{equation}
  \label{eq:gaugeon05}
  A_{\mu}\mapsto A_{\mu}+aD_{\mu}\varTheta,\quad
  \varPhi\mapsto\varPhi+aB,
\end{equation}
while disregarding the change of $A_{\mu}$ in $D_{\mu}\varTheta$.
Then we find that
$B\partial^{\mu}A_{\mu}-\varPhi\partial^{\mu}D_{\mu}\varTheta$ is
invariant under this $q$-number gauge transformation. (Again, the
similarity to the BRST invariance should be noted.) Therefore, if we
accept this restriction, our formulation of PIs developed in this
paper can be regarded as the gaugeon formalism for non-abelian gauge
theories. Despite the restriction stated above, it is useful for
treating gauge fields in perturbation theory. In this case, our PI
becomes that with the Lagrangian
\begin{equation}
  \label{eq:gaugeon06}
  {\mathcal L}_{\mathrm{I}}=
  -\frac{1}{4}F^{\mu\,\nu}F_{\mu\,\nu} +B\partial^{\mu}A_{\mu}
  -\varPhi\partial^{\mu}D_{\mu}\varTheta+
  \frac{\alpha}{2}\varPhi^{2}+
  i\bar{\eta}\partial^{\mu}D_{\mu}\eta+
  i\bar{c}\partial^{\mu}D_{\mu}c
\end{equation}
in the case of the type I formalism, and the Lagrangian
\begin{equation}
  \label{eq:gaugeon07}
  {\mathcal L}_{\mathrm{II}}=
  -\frac{1}{4}F^{\mu\,\nu}F_{\mu\,\nu} +B\partial^{\mu}A_{\mu}
  -\varPhi\partial^{\mu}D_{\mu}\varTheta+\frac{\alpha}{2}\varPhi B
  +i\bar{\eta}\partial^{\mu}D_{\mu}\eta
  +i\bar{c}\partial^{\mu}D_{\mu}c,
\end{equation}
in the case of the type II formalism, corresponding to those given in
Eqs.~\eqref{eq:thooft04} and \eqref{eq:typeII01}, respectively.

Turning now to the formulation for finite gauge transformations, we
consider the possibility of changing the gauge parameter through some
change of variables in our PIs. In a related context, Yokoyama, Takeda
and Monda~\cite{YTM} have formulated a gauge covariant canonical
quantization of non-abelian gauge theories. Its BRST symmetric version
was then constructed by Abe and also by Koseki, Sato and
Endo~\cite{gaugeonBRS2}. In their formulation, there exists a
parameter that can be changed under a $q$-number gauge transformation,
but this parameter cannot be identified with $\alpha$ in the linear
covariant gauge, $\partial^{\mu}A_{\mu}+\alpha B=0$. Due to this
discrepancy, the propagator of the gauge field becomes highly
complicated and different from that of the standard Lorentz-covariant
formulation. This sharply contrasts with the simplicity of 't~Hooft's
technique in the path integral formalism. To inquire further into this
matter would lead us into a specialized area that is irrelevant to the
main subject here, and such digression would undoubtedly obscure the
outline of our argument. We thus continue to examine the possibility
of formulating our PIs in a gauge covariant way.

Since, if we begin with the Landau gauge, as we have done throughout
this paper, the gauge parameter $\alpha$ enters the PI from the
Gaussian identity \eqref{eq:typeIidentity} [being equivalent to
Eq.~\eqref{eq:gaussiang01}] or from Eq.~\eqref{eq:typeIIidentity} for
the second type of gauge fixing, we may change $\alpha$ to an
arbitrary value by hand, using the $\alpha$-independence of these
formulae. However, we may also change $\alpha$ by means of some change
of variables in these identities. This can be done, as shown in
Appendix \ref{sec:appendixa} for the Gaussian identity, by solving the
equation $f[g[g{'},A],A]=\gamma f[g{'},A]$ for a given constant
$\gamma$. We therefore seek a change of variables from $g$ to $g{'}$
such that the scaling of $f[g,A]$ is generated. If the solution is
given by $g=g[g{'},A]$ as a functional of $g{'}$ and $A_{\mu}$, the PI
in terms of $g$ with the gauge parameter $\alpha$ is transformed to
that of $g{'}$ with $\alpha{'}=\alpha/\gamma^{2}$ for the case of
Eq.~\eqref{eq:typeIidentity} and $\alpha{'}=\alpha/\gamma$ for the
identity \eqref{eq:typeIIidentity}.

Although the change of variables from $g$ to $g{'}$ needed to satisfy
the scaling of $f[g,A]$ is quite complicated, we can confirm the
validity of our prescription as follows. Setting $f=-f[g^{-1},A]$, the
identities \eqref{eq:typeIidentity} and \eqref{eq:typeIIidentity}
are simplified as
\begin{equation}
  \label{eq:identities}
  1=\int\!\!{\mathcal D}f\,{\mathcal D}\varPhi\,
  {\mathcal D}\bar{\eta}\,{\mathcal D}\eta\,
  \exp\left(
    \int\!\!d^{4}x\,\tilde{\boldsymbol{\delta}}{'}\left[
      \bar{\eta}\left(f+\frac{\alpha}{2}V_{i}\right)\right]\right),\quad
  V_{\mathrm{I}}=\varPhi,\
  V_{\mathrm{II}}=B,
\end{equation}
in which the BRST transformation becomes
\begin{equation}
  \label{eq:brstsimple}
  \tilde{\boldsymbol{\delta}}{'}f=\eta,\
  \tilde{\boldsymbol{\delta}}{'}\bar{\eta}=i\varPhi,\
  \tilde{\boldsymbol{\delta}}{'}\eta=
  \tilde{\boldsymbol{\delta}}{'}V_{i}=0.
\end{equation}
Since these transformations and the measure of the integration
\eqref{eq:identities} are invariant under the scaling of these
variables\footnote{It seems that this was partly recognized by Koseki
  et al. in Ref.~\citen{gaugeonKSE}.} given by
\begin{equation}
  \label{eq:brstscale}
  f\mapsto e^{\rho}f,\
  \varPhi\mapsto e^{-\rho}\varPhi,\
  \bar{\eta}\mapsto e^{-\rho}\bar{\eta},\
  \eta\mapsto e^{\rho}\eta,
\end{equation}
in addition to $B\mapsto B$, we can rewrite Eq.~\eqref{eq:identities} as
\begin{equation}
  \label{eq:identities2}
  1=\int\!\!{\mathcal D}f\,{\mathcal D}\varPhi\,
  {\mathcal D}\bar{\eta}\,{\mathcal D}\eta\,
  \exp\left(
    \int\!\!d^{4}x\,\tilde{\boldsymbol{\delta}}{'}\left[
      \bar{\eta}\left(
        f+\frac{\alpha_{i}^{(\rho)}}{2}V_{i}\right)\right]\right),
\end{equation}
in which $\alpha_{\mathrm{I}}^{(\rho)}=e^{-2\rho}\alpha$ and
$\alpha_{\mathrm{II}}^{(\rho)}=e^{-\rho}\alpha$, corresponding to the
definition of $V_{i}$ above. Thus, we confirm the
$\alpha$-independence of these identities. (Though this was evident
from very beginning.) The important point here is that the BRST
transformation, given by Eq.~\eqref{eq:brstsimple}, commutes with the
scaling in Eq.~\eqref{eq:brstscale}. The change of variables from $g$
to $g{'}$ considered above for changing $\alpha$ is identical to the
scaling \eqref{eq:brstscale} when expressed in terms of $f$, $\varPhi$
and their ghosts. Furthermore, the BRST transformation
$\tilde{\boldsymbol{\delta}}{'}$ given above returns to that for $g$,
$\varPhi$ and their ghosts if we go back to the expression in terms of
these variables.  Hence, our method of changing $\alpha$ described
above also commutes with the BRST transformation. It is thus clear
that the structure of the total Hilbert space is preserved under such
a change of the gauge parameter $\alpha$. In view of these facts, we
conclude that the method presented here can be regarded as a gauge
covariant formulation that is useful even in the case of finite gauge
transformations for non-abelian theories.

\section{Summary}
\label{sec:summary}
We have proposed to regard the 't~Hooft average~\cite{tHooft} as a PI
over additional unphysical degrees of freedom. This allows us to
formulate a PI with extended gauge symmetry.~\cite{exgauge} \
Extension of the gauge invariance and the reduction obtained by
integrating the additional degrees out from the extended PI after the
gauge fixing for this extended gauge invariance are the keys to
understanding 't~Hooft's technique.  The remarkable feature of the
extended PI thus obtained is that it can be viewed as a hybrid of two
systems for different values of the gauge parameter of the original
degrees of freedom; different methods of integrating the additional
degrees of freedom, that is, different arrangements of the Lagrangian
and the gauge transformation of gauge fields in combination with
extended degrees of freedom, result in PIs for different gauges
of the original degrees of freedom.

We have also carried out analyses of the BRST invariance for the
extended system, finding as the symmetry of the extended system two
types of BRST transformations, that associated with the extended gauge
invariance and that resulting from the trivial nature of the systems
behind the identities useful for gauge fixing. This latter BRST
invariance can be regarded as a condition for specifying the Hilbert
space of the original degrees of freedom for a specific value of the
gauge parameter. Accordingly, by integrating the additional degrees of
freedom out of the extended PI, we observe that the total BRST charge
reduces to that of Kugo and Ojima~\cite{KO} and obtain the BRST
invariant Lagrangian of the original degrees of freedom for this
specific gauge.  In this sense, the BRST invariance of the extended
system can be regarded as a proper extension of the BRST symmetry in
the original system. Such a structure of the total Hilbert space,
viewed in the light of the BRST invariance, is very similar to that
found by Koseki et al.  in Ref.~\citen{gaugeonKSE} for the BRST
invariant formulation of Yokoyama's gaugeon formalism.~\cite{gaugeon}
\ In accordance with the hybrid structure of the extended PI, the
Hilbert space of the extended system has the same property, and this
allows us to write the total BRST charge in two ways.

The observation that a PI formulated by means of a gauge fixing
procedure from a gauge invariant one can be rewritten as a product of
a PI of the original degrees of freedom in a specific gauge with
another PI of the additional degrees of freedom which represents an
identity [i.e., \eqref{eq:typeIidentity} or
\eqref{eq:typeIIidentity}], as a cancellation of bosonic and fermionic
functional determinants with each other, provides another way of
finding the gauge fixed PIs of the extended system. We can regard the
identity mentioned above as a generalization of similar one, utilized
to formulate a PI by Koseki et al. in Ref.~\citen{gaugeonKSE}, to
non-abelian theories.  Furthermore, the use of such an identity plays
the central role in formulating a PI with extended degrees without
averaging over any $c$-number function.  (Avoiding the average over an
unknown $c$-number function is important when we consider the
canonical quantization of the system, as there exists no nice
prescription for understanding such a procedure in the operator
formalism.)  It is noteworthy that there exist systems with BRST
invariance underlying 't~Hooft's Gaussian identity and also in a
trivial relation from the functional integration of a delta function.
With this observation, we have noted that setting the $c$-number
function $f$ in the 't~Hooft average to
$f=-i\partial^{\mu}(gD_{\mu}g^{-1})$ can be regarded as a Nicolai
map.~\cite{nicolai} \ We then recognize that the systems possessing
these identities are the non-abelian counterparts of the Froissart
model\cite{froissart} with the BRST symmetry discussed in detail by
Kashiwa in Ref.~\citen{brstgauss}.

The invariance under the scaling of the variables, which commutes with
the BRST transformation, in the identities \eqref{eq:typeIidentity}
and \eqref{eq:typeIIidentity} allows us to change the gauge parameter
with this change of variables. Taking this fact into account, in
conjunction with the hybrid structure seen in the PIs, and also in the
total Hilbert space of the extended system, we recognize that the
total Hilbert space involves all linear covariant gauges of the
original system.  Since we can move freely in this total Hilbert space
to change the resulting gauge parameter, we conclude that the 't~Hooft
average, viewed from the point of view of our formulation of PIs with
extended degrees of freedom, is a generalization of Yokoyama's gaugeon
formalism to non-abelian gauge theories.

\section*{Acknowledgements}
The author would like to thank Professor T. Kashiwa for fruitful
discussions. He is also grateful to a referee for many valuable
suggestions. In particular, the analyses of the BRST invariance of the
extended system could not have been completed without the referee's advice.

\appendix
\section{BRST Invariant Formulation of a Gaussian Identity}
\label{sec:appendixa}
In this appendix we show that we can find a BRST invariance in a
Gaussian identity. Although a thorough explanation of the relation
between a Gaussian identity and the BRST invariance can be found in
Ref.~\citen{brstgauss}, here we present our own description, which is
useful for understanding presented in the main text, in particular the
additional BRST invariance of the PI in Eq.~\eqref{eq:thooft04}.

Let us consider the identity
\begin{equation}
  \label{eq:brstgauss01}
  1=\left(\frac{i}{2\pi\alpha}\right)^{n/2}
  \int\!\!d^{n}\varphi\,\exp\left(
    -\frac{i}{2\alpha}\varphi^{2}\right),\quad
  \varphi^{2}=\sum_{a=1}^{n}\varphi_{a}^{2}.
\end{equation}
If we regard $\varphi$ as a set of functions of $n$ independent
variables $x$, this can be rewritten as
\begin{equation}
  \label{eq:brstgauss02}
  1=\left(\frac{i}{2\pi\alpha}\right)^{n/2}
  \int\!\!d^{n}x\,\left\vert\frac{\partial\varphi}{\partial
      x}\right\vert\,
  \exp\left(
    -\frac{i}{2\alpha}\varphi^{2}(x)\right),
\end{equation}
where $\vert\partial\varphi/\partial x\vert$ is the Jacobian, assumed
to be positive definite hereafter, of the change of variables through
$\varphi=\varphi(x)$.  Apparently, this integral is invariant under
the change of variables $x\mapsto x{'}=x{'}(x)$, though the integrand
undergoes a change of functional form through
$\varphi(x(x{'}))=\varphi{'}(x{'})$ and the Jacobian $\vert\partial
x/\partial x{'}\vert$.  This invariance can be seen as the BRST
symmetry of the exponent in the integrand of
\begin{equation}
  \label{eq:brstgauss03}
  1=\int\!\!\frac{d^{n}x\,d^{n}k}{(2\pi)^{n}}\,
  (d\bar{c}\,dc)^{n}\,
  \exp\left(ik\varphi(x)+
    \frac{i\alpha}{2}k^{2}
    -\bar{c}D(x)c\right),\quad
  D(x)=\frac{\partial\varphi}{\partial x},
\end{equation}
where $c$ and $\bar{c}$ are a set of Grassmann variables, under
\begin{equation}
  \label{eq:brstgauss04}
  x\mapsto x{'}=x+\lambda c,\
  \bar{c}\mapsto \bar{c}{'}=\bar{c}+i\lambda k,\
  k\mapsto k,\ c\mapsto c
\end{equation}
with $\lambda$ a Grassmann parameter. If we set
$\boldsymbol{\delta}x=c$ and $\boldsymbol{\delta}\bar{c}=ik$ in the
above definition of the BRST transformation, we can express the
trivial nature of the Gaussian identity as
\begin{equation}
  \label{eq:brstgaussbrs}
  1=\int\!\!\frac{d^{n}x\,d^{n}k}{(2\pi)^{n}}\,
  (d\bar{c}\,dc)^{n}\,
  \exp\left(\boldsymbol{\delta}\left[\bar{c}\left\{
        \varphi(x)+\frac{\alpha}{2}k\right\}\right]\right).
\end{equation}
This is identical to the BRST invariance underlying the original
Gaussian identity \eqref{eq:brstgauss01}.

Because the right-hand side of Eq.~\eqref{eq:brstgauss01} is actually
independent of $\alpha$, we can change its value by hand to
$\alpha{'}$ without affecting the above argument. It is useful,
however, to note that the change in $\alpha$ can be generated by a
change of variables. To see this, let us suppose that $x$ is connected
to the new variable $y$ through the relation $x=x(y)$, so that
$\varphi(x(y))=\gamma\varphi(y)$ holds for a constant $\gamma$. Then
scaling $k$ as $\gamma k\mapsto k$, we find
\begin{equation}
  \label{eq:brstgauss05}
  k\varphi(x)+
  \frac{\alpha}{2}k^{2}
  \mapsto
  k\varphi(y)+
  \frac{\alpha}{2\gamma^{2}}k^{2}.
\end{equation}
Hence, we can carry out the change
$\alpha\mapsto\alpha{'}=\alpha/\gamma^{2}$ through this change of
variables, though it is equivalent, as explained above, to replacing
$\alpha$ with $\alpha{'}$ in Eq.~\eqref{eq:brstgauss01} by hand.

Let us now consider the Gaussian identity \eqref{eq:gaussiang} from
the point of view of the argument above. As is now clear, the original
form given by Eq.~\eqref{eq:gaussian} corresponds to the identity
\eqref{eq:brstgauss01}, and setting
$f=f[g,A]=-i\partial^{\mu}(gD_{\mu}g^{-1})$ is interpreted as the
analogue of $\varphi=\varphi(x)$ above. By exponentiating the Jacobian
$\varDelta[A^{g}]$ in terms of a fermionic PI, we obtain
\begin{equation}
  \label{eq:brstgauss06}
  1=\int\!\!{\mathcal D}g\,{\mathcal D}\varPhi\,
  {\mathcal D}\bar{\eta}\,{\mathcal D}\eta\,
  \exp\left(i
    \int\!\!d^{4}x\,\left\{\varPhi
      f[g,A]+\frac{\alpha}{2}\varPhi^{2}
      +i\bar{\eta}\partial^{\mu}D_{\mu}^{g}\eta^{g}\right\}\right),
\end{equation}
where $\eta^{g}=g\eta g^{-1}$, and the Fourier transform of the
Gaussian weight has also been carried out. If we parametrize $g$ as
$g=e^{i\varTheta}$, we find that the quadratic part of the Lagrangian
in this PI is given by
\begin{equation}
  \label{eq:brstgauss07}
  {\mathcal L}^{0}_{\mathrm{BRST}}=-\varPhi\partial^{\mu}\partial_{\mu}\varTheta+
  \frac{\alpha}{2}\varPhi^{2}+
  i\bar{\eta}\partial^{\mu}\partial_{\mu}\eta.
\end{equation}
Because this Lagrangian is simply the BRST invariant version of the
Froissart model,~\cite{froissart} \ we recognize that the Lagrangian
in the PI \eqref{eq:brstgauss06} can be understood as the non-abelian
generalization of the system discussed in the appendix of
Ref.~\citen{brstgauss}.  Then, the change of variables from $f$ to $g$
introduced in \S~\ref{sec:thooft} can be regarded as the Nicolai
map~\cite{nicolai} corresponding to the trivial nature of this system.
As a consequence, the system is invariant under the BRST
transformation
\begin{equation}
  \label{eq:brstgauss08}
  -ig\delta g^{-1}=\lambda g\eta g^{-1}
  =\lambda \eta^{g},\quad  
  \delta\bar{\eta}=i\lambda \varPhi,\quad
  \delta\eta=i\lambda \eta^{2}.
\end{equation}
By setting $\delta=\lambda\boldsymbol{\delta}$ in
Eq.~\eqref{eq:brstgauss08}, we can rewrite \eqref{eq:brstgauss06} as
\begin{equation}
  \label{eq:brstgauss09}
  1=\int\!\!{\mathcal D}g\,{\mathcal D}\varPhi\,
  {\mathcal D}\bar{\eta}\,{\mathcal D}\eta\,
  \exp\left(
    \int\!\!d^{4}x\,\boldsymbol{\delta}\left[\bar{\eta}
      \left(f[g,A]+\frac{\alpha}{2}\varPhi\right)\right]\right).
\end{equation}
This explains the trivial nature of the Gaussian identity
\eqref{eq:gaussiang} in terms of the BRST invariance. Finally, we note
that we can scale the gauge parameter $\alpha$ by transforming from
$g$ to $g{'}$, so that $f[g[g{'}],A]=\gamma f[g{'},A]$, as done above
in Eq.~\eqref{eq:brstgauss05}.

\end{document}